\documentclass[preprint,12pt]{elsarticle}

\usepackage{amssymb}
\usepackage{amsmath}
\usepackage{colortbl}
\usepackage{multicol}
\usepackage{rotating}
\usepackage{color}
\usepackage{graphicx}

\journal{Journal}
\begin{document}
\begin{frontmatter}



\title{Entropy-based link prediction in weighted networks}

\author[]{Zhongqi Xu}
\author[]{Cunlai Pu\corref{cor1}}
 \ead{pucunlai@njust.edu.cn}
 \cortext[cor1]{200 Xiaolingwei, Nanjing 210094, China. Tel: +8613915966537.}
 \author[]{Rajput Ramiz Sharafat}
\author[]{Lunbo Li}
\author[]{Jian Yang}

\address{Department of Computer Science and Engineering, Nanjing University of Science and Technology, Nanjing 210094, China}

\begin{abstract}
Information entropy has been proved to be an effective tool to quantify the structural importance of complex networks. In the  previous work (Xu et al, 2016 \cite{xu2016}), we measure the contribution of a path in link prediction with information entropy. In this paper, we further quantify the contribution of a path with both path entropy and path weight, and propose a weighted prediction index based on the contributions of paths, namely Weighted Path Entropy (WPE), to improve the prediction accuracy in weighted networks. Empirical experiments on six weighted real-world networks show that WPE achieves higher prediction accuracy than three typical weighted indices.
\end{abstract}

\begin{keyword}
Link prediction \sep Weighted networks \sep Information entropy
\PACS 89.75.Hc \sep 89.75.Fb \sep 89.20.Hh

\end{keyword}

\end{frontmatter}


\section{Introduction}
In the field of network science, real-world complex systems are abstracted as complex networks, in which nodes represent individuals and links denote the connections or interactions between individuals \cite{barabasi2016,newman2010,cheng2012}. Nowadays, although we can obtain abundant data of various complex systems due to advanced technologies, it is demonstrated that larger parts of the data of the complex systems are still not available, and there are non-ignorable errors in the data that we obtain \cite{mayer2013,abbasi2016}. Thus, new methods are needed to process, correct, and make predictions from the data. Link prediction methods aim to predict the missing or future links among network data \cite{lu2011,wang2015}. Specifically, they estimate the existence likelihood of links between two nodes based on observed links and nodes' attributes. Link prediction has broad applications \cite{lu2012}. For instance, it can be used in detecting potential interactions in protein-protein interaction network \cite{lei2013}, recommending friends and goods in online social networks \cite{sherkat2015}, exploring potential coauthor relationships in collaboration networks \cite{newman2001a} and so on.

Previous algorithms are basically from the field of machine learning including supervised learning \cite{al2006}, Markov Chain \cite{sarukkai2000}, and likelihood estimation \cite{getoor2005}. These algorithms heavily depend on attributes of nodes, and they do not seriously consider structural characteristics of networks. Besides, their computation cost is inhibitive for large real-world networks \cite{cui2016}. Recently, the booming network science community gets deeper insights into the structure of complex networks \cite{barabasi2016}, and further stimulates the research of link prediction \cite{lu2011}. Lots of prediction algorithms based on structural similarity are proposed, which can be classified into three types: local indices \cite{barabasi1999,newman2001b,kossinets2006,liben2007,addmic2003,zhou2009}, quasi-local indices \cite{lu2009,liu2010} and global indices \cite{katz1953,leicht2006}. For example, Common Neighbors (CN) \cite{newman2001b}, Preferential attachment (PA) \cite{barabasi1999}, Adamic-Adar (AA) \cite{addmic2003}, resource allocation (RA) \cite{zhou2009}, etc. are local indices, which only use the nearest neighborhood information. Katz \cite{katz1953}, Leicht-Holme-Newman (LHN) \cite{leicht2006}, SimRank \cite{glen2002} and so on, which need the knowledge of the whole network topology, are global ones. Quasi-local indices, including local path (LP) \cite{lu2009}, local random walk (LRW) \cite{liu2010}, Superposed Random Walk (SRW) \cite{liu2010}, etc. need more topological information than local indices, but less topological information than global ones. Generally, local indices have the lowest prediction accuracy, but their computational cost is the smallest among the three types of similarity indices. Global indices are the opposite of local indices, while quasi-local ones are the trade-off between them.

 Recently, information entropy has been employed to measure the complexity of the topological structures of complex networks \cite{sole2004}. Results showed that information entropy can better capture the topological difference than the other typical network measurements \cite{halu2014,anand2009}. Moreover, information entropy has nature connection with link prediction problem in that the probability of a missing link between two nodes can be transformed into the corresponding information entropy. Thus, researchers began to apply information entropy theory to link prediction problems in complex networks \cite{tan2014,xu2016,zhu2016}. For example, Tan et al \cite{tan2014} reexamined the role of common neighbors in link prediction by using the mutual information, and proposed a mutual information-based similarity index. Xu et al \cite{xu2016} derived the information entropy of a path, and studied the contributions of paths in link prediction based on path entropy, and finally provided a path entropy based similarity index. Simulation results \cite{tan2014,xu2016,zhu2016} showed that the similarity indices based on information entropy have higher prediction accuracy than the other types of similarity indices.

Many complex networks contain the information of link weights \cite{newman2010}, measuring the strength of connections between nodes, thus, it is reasonable to consider the link weights when designing link prediction algorithms to further improve the prediction accuracy. So far there have been a few tries in literature. Murata and Moriyasu \cite{murata2007} improved the CN and AA indices by using the link weights information. Bai Meng etc \cite{meng2011} developed the weighted version of LP index. L\"{u} L Y etc. \cite{lu2010} particularly explored the role of weak ties in link prediction and found that weak ties can improve the prediction accuracy effectively. Simulation results showed that the weighted version of these typical similarity indices indeed have a higher accuracy than the original ones. However,  they still have a large space to improve. In this paper, we apply information entropy to link prediction in weighted networks. Specifically, we reexamine the role of path in link prediction by considering both the path entropy and the path weight, and finally we propose a weighted similarity index. Through simulation on weighted real-world networks we show the prediction accuracy of our index and make comparison with other typical weighted indices.

\section{Path entropy}
Quantitative measurement of complex networks is a hot topic in network science \cite{barabasi2016}, and various measurements are proposed including degree, betweenness, closeness, K-Core number and so on \cite{newman2010}. However, most of the measurements are for nodes and links, and only a very few are particularly for paths such as path length \cite{pu2015a} and path attack centrality \cite{pu2015b}. In our latest work \cite{xu2016}, we applied information theory to measure the importance of paths, and specifically we derived the entropy of a path. Assuming that $L_{ab}^1$ ($L_{ab}^0$) represents the event  there is (not) a link between node $a$ and node $b$ in the network without degree correlation. Then, the probability of $L_{ab}^1$ is calculated as follows:
\begin{equation}
P(L^1_{ab})=1-P(L^0_{ab})=1-\prod_{i=1}^{k_b}{\frac{(M-{k_a})-i+1}{M-i+1}}=1-\frac{C_{M-{k_a}}^{k_b}}{C_M^{k_b}},
\end{equation}
where $k_a$ ($k_b$) is the degree of $a$ ($b$), and $M$ is the total link number in the network. Based on  the definition of entropy and Eq. (1), we  obtain the entropy of  $L_{ab}^1$  as follows:
\begin{equation}
I(L^1_{ab})=-\log(P(L^1_{ab}))=-\log (1-\frac{C_{M-{k_a}}^{k_b}}{C_M^{k_b}}).
\end{equation}

 Let's further consider a simple path $D=v_0v_1 \dots v_{\delta}$ of length $\delta$. The occurrence possibility of $D$ can be calculated approximately as:
\begin{eqnarray}
    P(D)\approx\prod_{i=0}^{\delta-1}{P(L^1_{v_iv_{i+1}})},
\end{eqnarray}
which indicates that the occurrence probability of a  path is approximately equal to the product of its links' occurrence probabilities. Thus, the entropy of path $D$ can be calculated  as\cite{xu2016}:
\begin{eqnarray}
I(D)=-\log(P(D))\approx\sum_{i=0}^{\delta-1}I(L^1_{v_iv_{i+1}})
\approx-\log(\prod_{i=0}^{\delta-1}(1-\frac{C_{M-{k_{v_{i+1}}}}^{k_{v_i}}}{C_M^{k_{v_i}}})).
\end{eqnarray}
Eq. (4) indicates that the entropy of a path approximates to the sum of its links' entropies. Furthermore, path entropy takes both path length and node degree information  into consideration. The longer path or the smaller node degree, the larger path entropy. Through the basic definition of entropy, we know that if a path has a large path entropy, then its occurrence probability should be low, in other words, the existence of the path is important to the network. Generally, path entropy  is more discriminating   than  the other measurements. The reason is that when computing the path entropy, we not only consider the path length, but also node degrees, and even the order of the nodes in the path.

\section{Path weight}
In real society, the interactions between individuals are of different strengths for many complex networks, which are called weighted networks \cite{newman2010}. For instance, in the air traffic network, the weight of a link is measured as the number of passengers in the related flight. In the router-level of the Internet, the weights of links are generally correlated with the bandwidth of the physical connections or the cost for data transmission between routers. In social networks, weights of links are related to the interacting times or frequencies between individuals. However, there is no specific research about the weights of paths in networks. In traffic routing protocols, we usually choose the optimal path, of which the sum of the cost of all links is the smallest among the candidate paths \cite{feigenbaum2005}. Enlightened by this, here we define the weight of a path $D$, as the sum of its links' weights, which is:
\begin{eqnarray}
    W_D=\sum_{t=0}^{\delta-1}{W_{{v_t}{v_{t+1}}}}.
\end{eqnarray}
where $ W_{{v_t}v_{t+1}}$ is the weight of the link with end nodes $v_t$ and $v_{t+1}$.

\section{Prediction index based on path entropy and path weight}
In the framework of information theory, the probability of link existence between two nodes can be expressed with information entropy. Then, the link prediction problem can be defined as the conditional entropy, which is as follows \cite{tan2014} :
\begin{equation}
I(L^1_{ab}|G')=I(L^1_{ab})-I(L^1_{ab};G'),
\end{equation}
where $G'$ is the topological structure information we know, and based on which we make the prediction of the event $L^1_{ab}$.  $I(L^1_{ab};G')$ is the joint entropy, which quantitatively measures how much the existence of $G'$ leads to the decrease of uncertainty of the event $L^1_{ab}$.
 In this paper, we consider the contributions of  simple paths in link prediction. Thus, we have $G'=\bigcup\limits_{i=2}^l{\{D_{ab}^i}\}$, where $\{D^i_{ab}\}$ is the set of all simple paths of length $i$ between  $a$ and $b$, and $l$ is the maximum length of simple paths we consider in the network.

 Previous results showed that the longer the path, the less important the paths in link prediction. Moreover, results demonstrated that the link weights can be used to improve the prediction accuracy. Here we combine the path length, path weight, and path entropy to calculate the contribution of  a path $D$ with length $i$ in the link prediction:
 \begin{eqnarray}
I(L_{ab}^1;D)\approx \frac{I(D)\cdot {W_{D}}^\alpha}{i-1},
\end{eqnarray}
where $\alpha$ is a free parameter, which controls the influence of path weights. When $\alpha=0$, path weights are ignored. When $\alpha>0$, path with large  weight is thought to have large contribution, otherwise, when $\alpha<0$, path with small weight is thought to have large contribution. $1/(i-1)$ is the optimal penalty factor that we found for suppressing the contributions of long paths.
Then, we give the definition of our prediction index, namely weighted path entropy (WPE) by considering the contributions of all simple paths, which is as follows:
\begin{eqnarray}
S_{ab}^{WPE}&=&I(L^1_{ab}|\bigcup_{i=2}^l{\{D_{ab}^i\}})\nonumber \\
           &=&I(L_{ab}^1)-I(L_{ab}^1;\bigcup_{i=2}^l{\{D_{ab}^i}\}) \nonumber \\
           &=&I(L_{ab}^1)-\sum_{i=2}^l{\{\frac{1}{i-1}(\sum_{D\in\{D_{ab}^i\}}\{{W_D}^{\alpha}\cdot I(D)\})\}}.
\end{eqnarray}
Based on Eq. (2), (4), (5) and (8), we finally obtain the complete expression of WPE index as follows:

\begin{eqnarray}
\resizebox{.9\hsize}{!}{$
S_{ab}^{WPE}=\log(\frac{C_M^{k_a}}{C_M^{k_a}-C_{M-{k_b}}^{k_a}})-\sum_{i=2}^l{\{\frac{1}{i-1}\sum_{D\in\{D_{ab}^i\}}\{\{{\sum_{t=0}^{i-1}{W_{{v_t}{v_{t+1}}}}}\}^{\alpha}\cdot \sum_{j=0}^{i-1}{\log(\frac{C_M^{k_{v_j}}}{C_M^{k_{v_j}}-C_{M-{k_{v_{j+1}}}}^{k_{v_j}}})\}}\}}.
$}
\end{eqnarray}

\section{Problem description and standard metrics}
 Assuming an undirected network $G(V,E,W)$, where $V$, $E$, and $W$ denote the sets of nodes, links and link weights respectively.  Note that in $W$, $W_{xy}=W_{yx}$, which means there is no direction for the weight between two nodes. To measure the predicting ability of an index, $E$ is usually randomly divided into two parts: a training set $E^T$ (in this paper, $90\%$ of all links) and a probe set $E^P$($10\%$ of all links). Clearly, $E=E^T\cup E^P$ and $E^T\cap E^P=\emptyset$.

 Two standard metrics AUC and Precision are applied to quantify the prediction quality. To calculate AUC,  $n$ times of independent score comparisons are made between node pairs in $E^P$ and  $U-E$. Each time we select a node pair randomly from these two sets respectively and compare their scores. If there are $n'$ times that the score of the link from $E^P$ is higher (or smaller in the case of WPE) than the link from $U-E$, and $n''$ times that they have the same scores, then, AUC is calculated as:
 \begin{equation}
AUC=\frac{n'+0.5n''}{n}.
\end{equation}
AUC should be about $0.5$ if all the scores are generated from an independent and identical distribution.
Precision aims to measure the ability to predict top ranked links. If among the top $L$ links ranked by the scores, there are $m $ links belonging to $E^P$, then Precision is calculated as:
 \begin{equation}
Precision=\frac{m}{L}.
\end{equation}
Note that for the WPE index, node pairs with small  scores ($S^{WPE}$ ) have large probability to be connected by  links. Thus, when calculating the Precision, we rank the links based on their $S^{WPE}$ from the smallest to the largest, and thus the top ranked links have the smallest scores,  which is the opposite to the other prediction indices.

Next, we introduce three indices and their weighted versions that we use for comparison. They are Common neighbors (CN) \cite{newman2001b}, Adamic-Adar Index (AA) \cite{addmic2003} and Local Path (LP) \cite{lu2009}, which are defined as follows:
 \begin{equation}
S^{CN}_{ab}=|O_{ab}|,
\end{equation}
\begin{equation}
S^{AA}_{ab}=\sum_{c\in{O_{ab}}}{\frac{1}{\log(|\Gamma(c)|)}},
\end{equation}
\begin{equation}
S^{LP}=A^2+\epsilon A^3,
\end{equation}
where $O_{ab}$ is the set of common neighbors of node $a$ and $b$,  and $\Gamma(c)$ is the set of neighbors of node $c$. $A$ is the adjacency matrix, and we set $\epsilon=0.01$ to obtain a near optimal prediction accuracy. In addition, their parameter-dependent weighted versions, WCN \cite{murata2007}, WAA \cite{murata2007}, and WLP \cite{meng2011} are respectively defined as follows:
 \begin{equation}
S^{WCN}_{ab}=\sum_{c\in{O_{ab}}}{{(W_{xc}}^\alpha+{W_{cy}}^\alpha)},
\end{equation}
 \begin{equation}
S^{WAA}_{ab}=\sum_{c\in{O_{ab}}}{\frac{{(W_{xc}}^\alpha+{W_{cy}}^\alpha)}{\log(1+S_c)}},
\end{equation}
 \begin{equation}
S^{WLP}_{ab}=\sum_{c\in{O_{ab}}}{{(W_{xc}}^\alpha+{W_{cy}}^\alpha)}+\epsilon \sum_{{(i,j)\in l_{x\to y}}}{{(W_{xi}}^\alpha+{W_{ij}}^\alpha)({W_{ij}}^\alpha+{W_{jy}}^\alpha)},
\end{equation}
where $S_c=\sum_{z\in\Gamma(c)}{{W_{cz}}^\alpha}$. $l_{x\to y}$ is the path of length three between  node $x$ and node $y$.  $i$ and $j$ are the intermediate nodes in the path $l_{x\to y}$. CN and AA as well as their weighted versions are all taken as  local indices which are only based on the nearest neighbors.  LP and its weighted version are quasi-local indices since they consider the next-nearest neighbors.

\section{Results}
Six weighted networks from disparate fields are used to test the  accuracy for various prediction indices. The directions of links are ignored.  The self-connections and multiple links are deleted from the network data. For the unconnected networks, we select the maximum connected components for experiments.  The statistics of these networks are summarized in Table 1. $\romannumeral1)$ Les \cite{les}:  This undirected network contains co-appearances of characters in Victor Hugo's novel `Les Mis\'{e}rables'.
A node represents a character and a link between two nodes shows that these two characters appeared in the same chapter of the book. The weight of each link indicates how often such a co-appearance occurred.
$\romannumeral2)$ USAir \cite{usair}: The network of US air transportation, where nodes represent airports, links represent routes between airports, and the weight of a link is the frequency of flights between two airports.
$\romannumeral3)$ Bomb \cite{bomb}: This undirected network contains contacts between suspected terrorists involved in the train bombing of Madrid on March 11, 2004 as reconstructed from newspapers. A node represents a terrorist and a link between two terrorists shows that there was a contact between the two terrorists. The link weights denote how `strong' a connection was, by considering the friendship and co-participating in training camps or previous attacks.
$\romannumeral4)$ Bible \cite{bible}: This undirected network contains nouns (places and names) of the King James Version of the Bible and information about their co-occurrences. A node represents one of the above noun types and a link indicates that two nouns appeared together in the same Bible verse. The link weights show how often two nouns occurred together.
$\romannumeral5)$ Florida \cite{florida}: This network contains the carbon exchanges in the cypress wetlands of South Florida during the dry season. Nodes represent taxa and a link denotes that a taxon uses another taxon as food with a given trophic factor (which is the basis of link weight).
$\romannumeral6)$ C.elegans \cite{celegans}: the neural network of the nematode worm C. elegans, where a node represents a neuron, a link joins two nodes if the corresponding neurons have synaptic contacts, and the weight represents the number of synapses between two neurons.

\begin{table}\centering
\begin{tabular}{cccccc}
\hline
Networks &$|V|$  &$|E|$  &$<k>$ &$H$  &$C$\\
\hline
Les&77&278&7.2338&1.9654&0.4634\\
USAir&332&2126&12.8072&3.4639&0.7494 \\
Bomb&70&256&7.3429&2.3645&0.7122 \\
Bible&1773&9131&10.3001&4.0115&0.7208 \\
Florida&128&2106&32.9063&1.2307&0.3346 \\
C.elegans&297&2148&14.4646&1.8008&0.3079 \\
\hline \\
\end{tabular}
\caption{The  topological statistics of six real-world weighted networks. $|v|$ is the number of nodes. $|E|$ represents the number of links. $<k>$ is the average degree.  $H$ represents the degree heterogeneity, defined as $H=\frac{<k^2>}{{<k>}^2}$. $C$ is the clustering coefficient.}
\end{table}

To investigate the ability of our prediction index, we perform experiments on the six weighted networks, and make comparison with three typical indices. Both the unweighted and weighted versions of these indices are tested. Note that the link weights of the networks are only considered when calculating the weighted versions of these indices. In addition, for the weighted versions of these indices, we adjust the control parameter $\alpha$ and achieve the near optimal performances measured by AUC and Precision. For the PE and WPE indices, $l=2$ means only  paths with length of 2  are used in the calculation, while $l=3$ indicates that paths with lengths of both 2 and 3 are used in the calculation. Our simulation results are shown in Table 2 and 3, as well as Fig. 1 and 2. Each value is the average of 100 independent runs. In the tables, the maximum performances for each network are marked in bold font.

\begin{table}
\resizebox{\textwidth}{!}{ %
\begin{tabular}{ccccccccccc}
\hline
Nets$\diagdown$index&$WAA$&$AA$&$WLP$&$LP$&$WCN$&$CN$&$WPE(l=2)$&$PE(l=2)$&$WPE(l=3)$&$PE(l=3)$ \\ \hline
Les&0.967477(0.6)&0.962911&0.96254(0.3)&0.955112&0.963042(0.3)&0.955129&\textbf{0.97055}(0.6)&0.964639&0.956291(0.4)&0.953417\\
USAir&0.967732(-0.5)&0.965757&0.954597(-0.3)&0.951873&0.963738(-0.7)&0.953757&\textbf{0.971917}(-0.2)&0.969813&0.966167(-1)&0.946596\\
Bomb&0.944603(0.7)&0.944998&0.938329(0.2)&0.937651&0.934543(0.5)&0.934408&\textbf{0.957373}(1.2)&0.954435&0.940593(-0.5)&0.939566\\
Bible&0.985164(0.2)&0.985157&0.978042(-0.8)&0.975619&0.977955(-0.1)&0.976989&\textbf{0.988501}(0.2)&0.987872&0.964088(-0.9)&0.95649\\
Florida&0.611155(0.1)&0.606072&0.783328(-0.3)&0.668619&0.609324(0.1)&0.604758&0.59479(1.4)&0.562404&\textbf{0.863607}(0.1)&0.860174\\
C.elegans&0.868117(0.1)&0.866639&0.868887(0)&0.868344&0.850465(-0.1)&0.850071&0.87321(0.2)&0.871474&\textbf{0.885356}(0)&0.885356\\
\hline \\
\end{tabular}}
\caption{Prediction accuracy measured by AUC for the unweighted and weighted  indices.  For the weighted indices, we present the maximum AUC with the corresponding control parameter $\alpha_{opt}$ in the brackets.}
\end{table}

\begin{table}
\resizebox{\textwidth}{!}{ %
\begin{tabular}{ccccccccccc}
\hline
Nets$\diagdown$index&$WAA$&$AA$&$WLP$&$LP$&$WCN$&$CN$&$WPE(l=2)$&$PE(l=2)$&$WPE(l=3)$&$PE(l=3)$ \\ \hline
Les&0.2212(0.2)&0.2192&0.2077(0.3)&0.2043&0.2162(0.3)&0.2112&\textbf{0.2227}(0.5)&0.2205&0.21(0.2)&0.2076\\
USAir&0.6514(-0.3)&0.6324&0.6207(-0.4)&0.6094&0.633(-0.3)&0.6115&0.5926(0.1)&0.5854&\textbf{0.6704}(-0.1)&0.6672\\
Bomb&0.2026(0)&0.1988&0.1749(-1)&0.1678&0.1886(0.5)&0.188&\textbf{0.2157}(1.6)&0.2101&0.1861(-0.9)&0.1814\\
Bible&0.5853(-1.5)&0.5602&0.3839(-0.4)&0.3798&0.4375(-0.1)&0.4371&\textbf{0.8536}(-0.1)&0.8508&0.3859(0.4)&0.3812\\
Florida&0.144(0.3)&0.0912&0.2797(-0.2)&0.1314&0.1384(0.2)&0.0965&0.0614(0.1)&0.0228&\textbf{0.4826}(0)&0.4826\\
C.elegans&0.1895(1.4)&0.1328&0.1757(0.6)&0.1371&0.1824(1.5)&0.1384&0.1293(0.2)&0.1262&\textbf{0.2044}(0.3)&0.1841\\
\hline \\
\end{tabular}}
\caption{Prediction accuracy measured by Precision (top-$100$) for the unweighted and weighted  indices.}
\end{table}

\begin{figure}
 \centering
\includegraphics[width=\textwidth,height=2.5in]{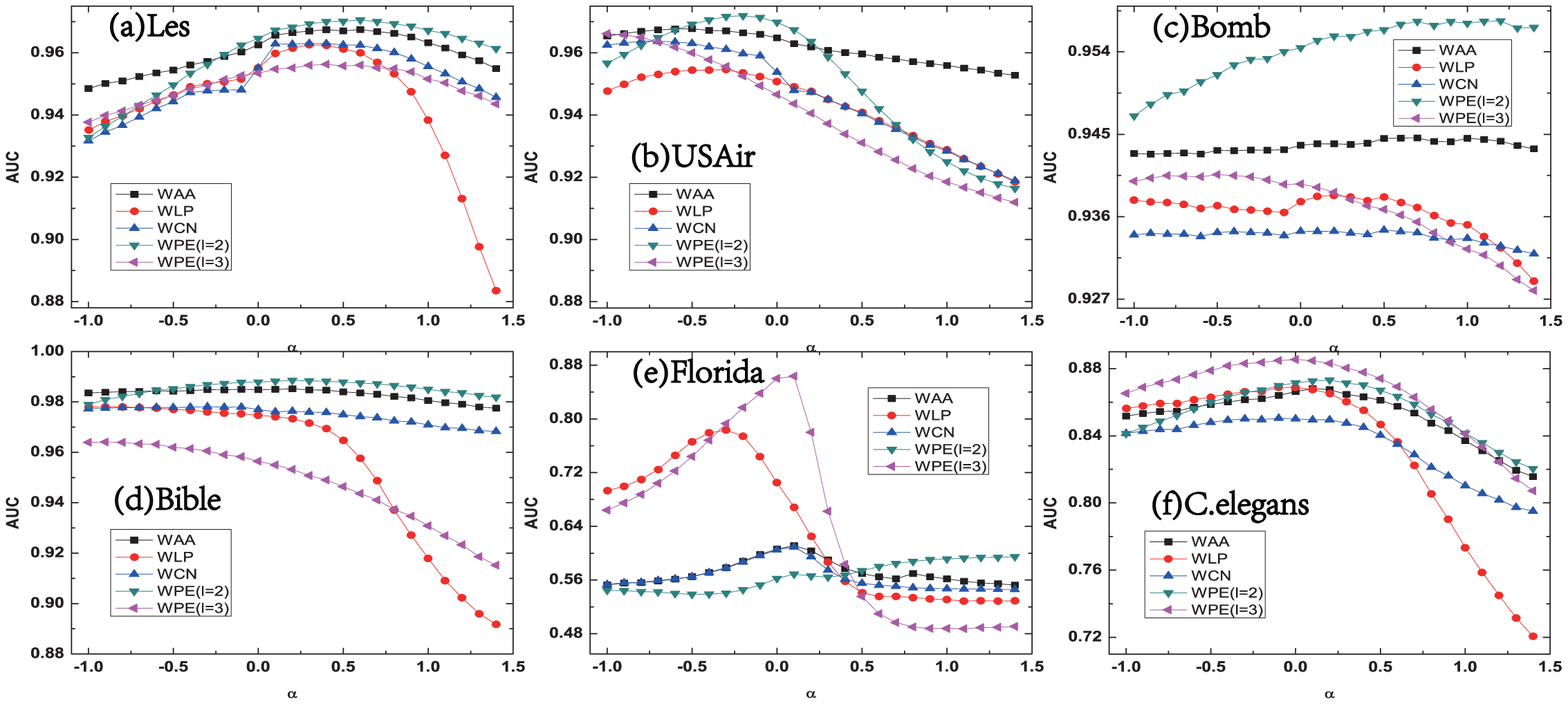}
\parbox[c]{15.0cm}{\footnotesize{\bf Fig.~1.} AUC vs. $\alpha$ for the weighted indices.}
\end{figure}

\begin{figure}
 \centering
\includegraphics[width=\textwidth,height=2.5in]{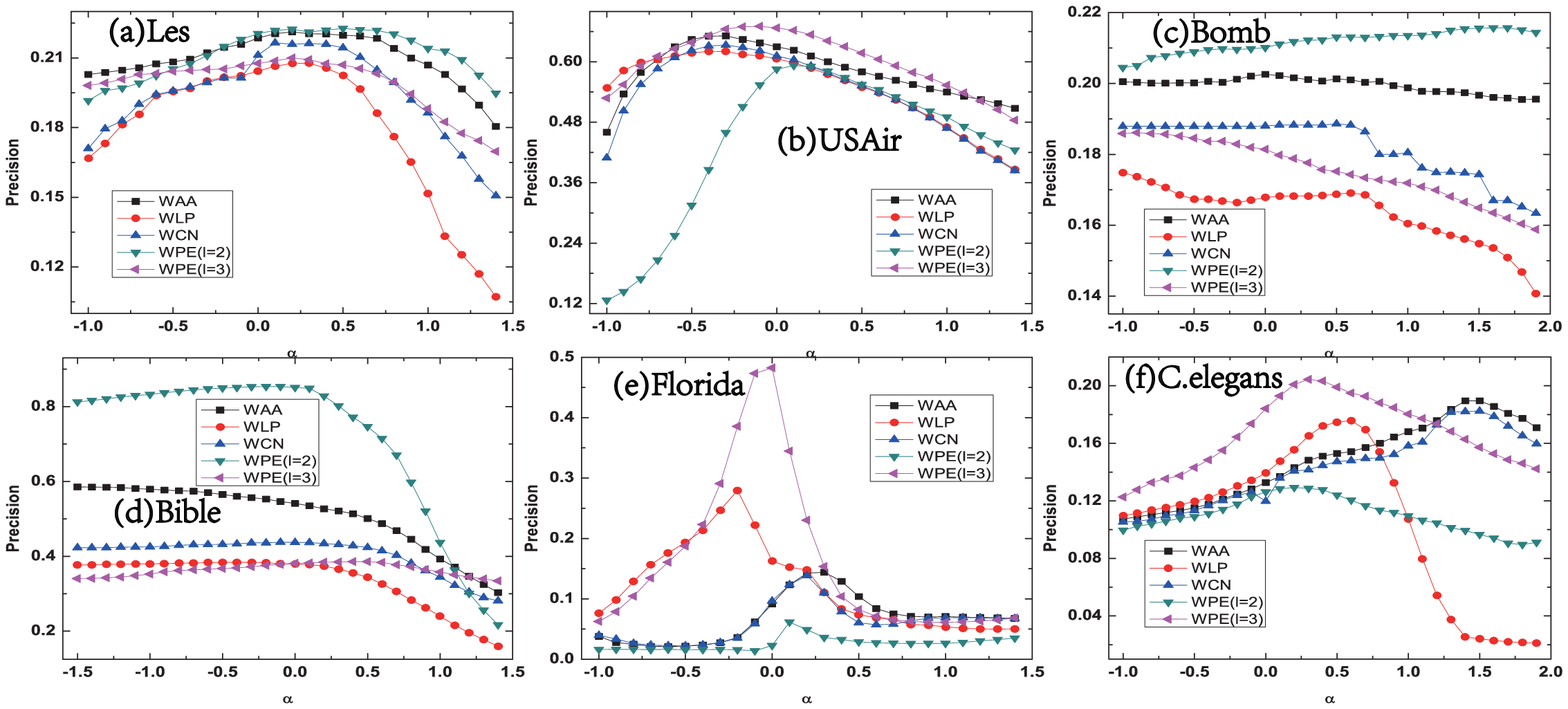}
\parbox[c]{15.0cm}{\footnotesize{\bf Fig.~2.} Precision (top-$100$) vs. $\alpha$ for the weighted indices.}
\end{figure}

Table 2 and Table 3 generally show that it is necessary to consider the link weights since the weighted version of these indices achieves higher prediction accuracy than the unweighted ones when the appropriate control parameter is considered (The near optimal parameters $\alpha_{opt}$ are given in the tables). Furthermore, Table 2 indicates that from the perspective of AUC, the prediction accuracy of all the unweighted indices is already high for all the six real-world networks, so that the improvements of the weighted versions are not significant. However, as shown in Table 3, the improvements of the weighted versions are more obvious from the perspective of Precision. Moreover, through Table 2 and 3 we see that for all the unweighted and weighted prediction indices, WPE always have the best performances for all the six real-world networks. Especially, for the networks of Bible, Florida, C. elegans and USAir, we can see from Table 3 that PE and WPE are much better than the other prediction indices. It is worth mentioning that for PE and WPE, extra consideration of the contributions of longer paths is not always necessary and sometimes negative for link prediction. For instance, from Table 2 and 3 we see that for some networks $l=2$ is better than $l=3$.

Fig. 1 and 2 show how the weight affects the prediction accuracy, for the weighted indices. Note that for WPE, we focus on path weight defined as the sum of all the link weights in the path, while the other weighted indices only consider link weights. From Fig. 1 and 2, we can see that  for WPE, generally both AUC and Precision increase first, and then decrease with the control  parameter $\alpha$, and there are optimal parameters $\alpha_{opt}$ (which are given in Table 2 and 3) corresponding to the maximum performances. The curves of the other weighted indices have the similar change trends. These results indicate that the prediction accuracy is sensitive to path weights for WPE and link weights for the others, and we should balance the influences of the weak ties and strong ties by appropriately choosing the control  parameter to achieve high prediction accuracy. Moreover, we should be slightly biased to the contributions of weak ties, since the simulations results demonstrate that the optimal parameters are less than 1 for all the six networks except Bomb (For WAA and WCN, C. elegans is also an exception).

\section{Conclusion}
In summary, we study the link prediction problem in weighted complex networks from the perspective of information entropy. In fact, the likelihood of a link between two nodes can be converted into entropy, and small entropy corresponds to large probability of link existence. In this paper we consider the contributions of simple paths between node pairs in link prediction. Specifically, we measure the contribution of an existing path by considering its  length,  entropy,  and  weight, which is further defined as the sum of link weights in the path. Furthermore,  we propose a weighted path entropy (WPE) index for link prediction by considering the contributions of all existing simple paths. Through simulation on several real-world weighted networks, we found that WPE has higher prediction ability measured by AUC and Precision than the other typical weighted similarity indices. In fact, the PE index proposed in our previous work already has high prediction ability. Thus, by appropriately considering the path weight, WPE further improves the prediction accuracy. Through simulation, we also found that weak tie is more critical than the strong tie in link prediction. Note that in our context, weak tie refers to path with small weight, while in the other works weak tie means small weight link.

\section*{Acknowledgments}
This work was  supported by the National Natural Science Foundation of China (Grant Nos. 61304154), the Specialized Research Fund for the Doctoral Program of Higher Education of China  (Grant No. 20133219120032),  the Postdoctoral Science Foundation of China (Grant No. 2013M541673), and China Postdoctoral Science Special Foundation (Grant No. 2015T80556).






\end{document}